\documentstyle[12pt,amsbsy]{article}
\newcommand{\beq}{\begin{equation}}
\newcommand{\eeq}{\end{equation}}

\begin{document}

\begin{titlepage}

\vspace{1cm}

\begin{center}
{\large Entropy and Area of Black Holes in Loop Quantum Gravity}
\end{center}

\begin{center}
I.B. Khriplovich\footnote{khriplovich@inp.nsk.su}
\end{center}
\begin{center}
Budker Institute of Nuclear Physics\\
630090 Novosibirsk, Russia,\\
and Novosibirsk University
\end{center}

\bigskip

\begin{abstract}
Simple arguments related to the entropy of black holes strongly
constrain the spectrum of the area operator for a Schwarzschild
black hole in loop quantum gravity. In particular, this spectrum
is fixed completely by the assumption that the black hole entropy
is maximum. Within the approach discussed, one arrives in loop
quantum gravity at a quantization rule with integer quantum
numbers $n$ for the entropy and area of a black hole.

\end{abstract}

\vspace{8cm}

\end{titlepage}

The quantization of black holes was proposed long ago in the
pioneering work \cite{bek}, and from other points of view in
\cite{muk, kog}. The idea of \cite{bek} was based on the
intriguing observation \cite{chr} that the horizon area $A$ of a
nonextremal black hole behaves in a sense as an adiabatic
invariant. This last fact makes natural the assumption that the
horizon area should be quantized. Once this hypothesis is
accepted, the general structure of the quantization condition for
large (generalized) quantum numbers $N$ gets obvious, up to an
overall numerical constant $\alpha$. The quantization rule should
be
\beq\label{An}
A_N=\alpha \,l_p^2 \,N.
\eeq
Indeed, the presence of the Planck length squared
\beq
l_p^2 = G\hbar /c^3
\eeq
in formula (\ref{An}) is only natural. Then, for $A$ to be finite
in a classical limit, the power of $N$ in expression (\ref{An})
should be equal to that of $\hbar$ in $l_p^2$. This argument,
formulated in \cite{khr}, can be checked, for instance, by
inspecting any expectation value nonvanishing in the classical
limit in ordinary quantum mechanics.

The subject of the present note is the entropy and spectrum of
black holes in loop quantum gravity [6-10]. We confine below to a
rather simplified version of this approach where the area spectrum
of a spherical surface is
\beq\label{Aj}
A =\alpha\, l_p^2 \sum_{i=1}^{\nu}\sqrt{j_i(j_i+1)}.
\eeq
Here a half-integer or integer ``angular momentum'' $j_i$,
\beq\label{j}
j_i=1/2, 1, 3/2, ...\; ,
\eeq
is ascribed to each of $\nu$ edges intersecting the surface. This
set of edges, labeled by index $i$, determines the surface
geometry. The quantum numbers $j_i$ assigned to these edges are
constrained by the condition
\beq\label{n}
\sum_{i=1}^{\nu} j_i = n,
\eeq
where $n$ is an integer. To each ``angular momentum'' ${\bf j}_i$
one ascribes $2j_i+1$ possible projections, from $-j_i$ to
$+j_i$. We assume below that for a given ${\bf j}_i$ all these
projections have the same weight. With the vectors ${\bf j}_i$
being the only building blocks of the model, it is natural to
consider that this $2j_i+1$ degeneracy for each angular momentum
${\bf j}_i$ corresponds to the spherical symmetry of the surface.

Some resemblance between expressions (\ref{An}) and (\ref{Aj}) is
obvious, though in the last case the large number
\beq\label{N}
N= \sum_{i=1}^{\nu}\sqrt{j_i(j_i+1)}
\eeq
is certainly no integer. As to the overall numerical factor
$\alpha$ in (\ref{Aj}), it cannot be determined without an
additional physical input. This ambiguity originates from a free
(so-called Immirzi) parameter \cite{imm,rot} which corresponds to
a family of inequivalent quantum theories, all of them being
viable without such an input. One may hope that the value of this
factor in (\ref{Aj}) can be determined by studying the entropy of
a black hole. This idea (mentioned previously in~\cite{boj}) is
investigated below.

We define the entropy $S$ of a spherical surface as the logarithm
of the number of states of this surface with fixed $n$, $\nu$, and
$\nu_j$, where $\nu_j$ is the number of edges with given $j$. Due
to the mentioned $2j+1$ degeneracy for each ``angular momentum''
$j$, the entropy is
\beq\label{en1}
S=\ln\left[\prod_j(2j+1)^{\nu_j}\frac{\nu\,!}{\prod_j\nu_j\,!}\right]
=\sum_j \nu_j \ln(2j+1) + \ln (\sum_j\nu_j\,!) - \sum_j \ln
\nu_j\,!.
\eeq
The obvious constraints are
\beq\label{con}
\sum_j \nu_j = \nu; \quad \sum_j j\,\nu_j = n.
\eeq

Let us mention that the entropy arguments exclude for a black hole
``empty'' edges with $j_i=0$. Obviously, if ``empty'' edges were
allowed, the entropy would be indefinite even for fixed $N$ and
$n$. In particular, with ``empty'' edges the Bekenstein-Hawking
relation
\beq\label{beha}
S=\,\frac{A}{4\,l_p^2}
\eeq
would not hold. Moreover, by adding an arbitrary number $\nu_0$ of
``empty'' edges in arbitrary order, the entropy could be made
arbitrarily large without changing $N$ and $n$. Indeed, with
``empty'' edges allowed, the ratio $\nu\,!/\nu_0\,!$ grows
indefinitely with $\nu_0$ at fixed values of $\nu_j$ with $j\neq
0$.

On the other hand, the same fundamental relation (\ref{beha})
dictates that the number of edges $\nu$ should be roughly on the
same order of magnitude as the sum $n$ of ``angular momenta''.
Let us mention in this connection the model proposed
in~\cite{boj}. In this model the horizon is characterized by a
single edge with $j=n/2$. Then the entropy grows with $n$
logarithmically,
\[
S=\ln(2j+1)=\ln(n+1) \to \ln n\, ,
\]
while the area grows with $n$ linearly,
\[
A  \sim \sqrt{j(j+1)} \sim \sqrt{n(n+2)}\; \to n\, .
\]
Since the requirement (\ref{beha}) for the classical limit $n \to
\infty$ is grossly violated in it, the model of~\cite{boj} has no
physical meaning, or at least is incomplete. To save the model,
some extra source of degeneracy should be included into it, but
one cannot find in~\cite{boj} any mention of such a degeneracy.

Thus, at least in the approach discussed (as distinct for example
from that of \cite{sm}), relation (\ref{beha}) is an absolutely
nontrivial constraint on a microscopic structure of theory.

It is natural to consider that the entropy of an eternal black
hole in equilibrium is maximum. This argument is emphasized in
\cite{vaz}, and used therein in a model of the quantum black hole
as originating from dust collapse. Just the discussion of the
assumption of maximum entropy is the main subject of the present
paper. More definite formulation of the problem considered below
is as follows. With the quantum numbers $j_i$ being the only
building blocks of the model, we are looking for such their
distribution over the edges which results in the maximum entropy
for a fixed total amount of the building material $n=\sum j_i$.

It is rather obvious intuitively that the entropy is maximum when
all values of $j$ are allowed. To demonstrate that this is
correct,  we will consider few more and more complex examples step
by step, starting with the simplest choice for the quantum numbers
$j_i$, where all of them are put equal to $1/2$. Then $\nu_j =\nu
\delta_{j,1/2}, \; \nu =2n$, and
\beq\label{s2}
S=2\ln2\,n.
\eeq
With all $j_i=1/2$ and $\nu =2n$, the area given by formula
(\ref{Aj}) equals
\beq\label{A1/2}
A =\alpha\, l_p^2\,\frac{\sqrt{3}}{2}\,\nu=\alpha\,
l_p^2\,\sqrt{3}\,n.
\eeq
Now, under the made assumption we obtain, due to formulae
(\ref{beha}) -- (\ref{A1/2}), the following value of the parameter
$\alpha$ of the theory:
\beq\label{al1/2}
\alpha=\,\frac{8\ln2}{\sqrt{3}}.
\eeq
It should be pointed out that this is the value of the parameter
$\alpha$ derived previously in \cite{asht} within a Chern-Simons
field theory, and that the typical value of $j_i$ obtained therein
is also $1/2$ (see also~\cite{kama,car}).

In fact, in this way one arrives at the quantization rule for the
black hole entropy (and area) with integer quantum numbers $\nu$
or $n$ (see formula (\ref{s2})), as proposed in \cite{bek}.
Moreover, in this picture the statistical weight of the quantum
state of a black hole is $2^\nu$ with integer $\nu$, as argued in
\cite{muk} (in the present case this integer $\nu$ should be
even).

Let us include now $j=1$ in line with $j=1/2$. Then the entropy
reaches its maximum value
\[
S=2\ln 3\, n = 2.197 \,n
\]
for $\nu_{1/2}=n, \; \nu_1=n/2$, with the mean value $<j>$ of
angular momenta
\[
<j>=n/\nu=2/3=0.667.
\]
(Here and below we retain in the expressions for entropy only
leading terms, linear in the large parameter $n$.) It is curious
to compare these numbers with the analogous ones $S=2\ln
2\,n=1.386 \,n$, $<j>=j=1/2=0.5$ for the pure $j=1/2$ case.

But what happens if quantum numbers larger than 1 are also
allowed? When $j=3/2$ are included, in line with $j=1/2$ and
$j=1$, the maximum entropy value
\[
S=2.378 \,n
\]
is attained at $\nu_{1/2}= 0.810\, n, \; \nu_1= 0.370\, n, \;
\nu_{3/2}= 0.150\, n $. Both entropy and average angular momentum
\[
<j>=0.752
\]
increase again, but not too much, as compared to the previous ones
$S=2.197\,n$, $<j>=0.667$.

It is natural now to expect that the absolute maximum of entropy
is reached when all values of quantum numbers are allowed. Let us
consider this situation starting with the general formula
(\ref{en1}). It is convenient to go over in it to new variables
$y_j$:
\beq
\nu_j=n y_j\,,
\eeq
constrained in virtue of (\ref{n}) by the obvious relation
\beq\label{con1}
\sum_j j \, y_j =1.
\eeq
Then, by means of the Stirling formula for factorials, we
transform (\ref{en1}) to the following expression:
\beq\label{en2}
S=n\left[\sum_j y_j \,\ln (2j+1) + \sum_j y_j \times \ln
(\sum_{j^{\prime}} y_{j^{\prime}}) - \sum_j y_j \ln  y_j \right].
\eeq
Only the contribution proportional to the large number $n$ is
retained here. We have assumed also that the number of essential
terms in the sums entering (\ref{en1}) (i. e. the number of the
essential classes of the edges with the same $j$) is much smaller
than $n$. In fact, this number is on the order of $\ln n$, and
the leading correction to the approximate formula (\ref{en2}) is
on the order of $\ln^2 n$. Again, the situation with the leading
correction here is different from that for the case when all
$j_i=1/2$, where the correction is just absent, and from that for
the model considered in \cite{kama,car}, where it is on the order
of $\ln n$.

We are looking for the extremum of expression (\ref{en2}) under
the condition (\ref{con1}). The problem reduces to the solution of
the system of equations
\beq\label{sys}
\ln (2j+1) +  \ln (\sum_{j^{\,\prime}} y_{j^{\,\prime}}) - \ln y_j
= \mu j,
\eeq
or
\beq\label{y}
y_j = (2j+1)\, e^{- \mu j}\,\sum_{j^{\,\prime}}
y_{j^{\,\prime}}\,.
\eeq
Here $\mu$ is the Lagrange multiplier for the constraining
relation (\ref{con1}). Summing expressions (\ref{y}) over $j$, we
arrive at equation
\beq
\sum_{j=1/2}^\infty (2j+1)\, e^{- \mu j}= 1, \quad \mbox{\rm or}
\quad \sum_{p=1}^\infty (p+1)z^p=1, \quad p=2j\,, \quad z=e^{-
\mu/2}\,.
\eeq
Its solution is readily obtained:
\beq\label{z}
z=1-\frac{1}{\sqrt{2}}\,,\quad \mbox{or} \quad \mu=-2\ln z =
2.456\,.
\eeq
Let us multiply now equation (\ref{sys}) by $y_j$ and sum over
$j$. Then, with the constraint (\ref{con1}) we arrive at the
following result for the absolute maximum of the entropy for a
given value of $n$:
\beq\label{enf}
S= \mu \,n= 2.456\, n\,.
\eeq
This is the final term of the succession of previous values of
entropy $S$:
\[
1.386 \,n\, \quad 2.197\, n, \quad 2.378 \,n\,.
\]
Assuming that the entropy of an eternal black hole in equilibrium
is maximum, we come to the conclusion that it is just
(\ref{enf}), which is the true value of the entropy of a black
holeî

To find the mean angular momentum $<j>$ in the state of maximum
entropy, let us rewrite the constraint (\ref{con1}) as
\beq
y_{1/2} \sum_{p=1}^\infty (p+1) p\, z^{p-1}=1.
\eeq
The sum in the last expression is also easily calculated, and
with the value (\ref{z}) for $z$ we obtain
$y_{1/2}=1/\sqrt{2}\,$. In its turn, this value of $y_{1/2}$
together with equation (\ref{y}) gives
\beq\label{yj}
y_j=\,\frac{1}{2\sqrt{2}}\,(2j+1)z^{2j-1},
\eeq
and
\beq\label{ys}
\sum_{j=1/2}^\infty y_j=\,\frac{\sqrt{2}+1}{2}\,=1.207.
\eeq

Now, the mean angular momentum is
\beq\label{<j>}
<j>=\frac{n}{\nu}=\left[\sum_{j=1/2}^\infty y_j\right]^{-1}
=2(\sqrt{2}-1)=0.828,
\eeq
which fits perfectly the succession of previous mean values $<j>$:
\[
0.5\, \quad 0.667\, \quad 0.752\,.
\]

Let us come back to the expression (\ref{Aj}) for the black hole
entropy. The sum (\ref{N}) is conveniently rewritten as
\[
N=\sum_{j=1/2}^\infty \sqrt{j(j+1)}\; \nu_j\,.
\]
With our formulae (\ref{z}), (\ref{yj}), one can easily express
this sum via $n$:
\[
N=1.471\,n.
\]
Thus, using the Bekenstein-Hawking relation (\ref{beha}), we
obtain the following results in the loop quantum gravity for the
area $A$ of an eternal spherically symmetric black hole in
equilibrium and for the constant $\alpha$ of the area spectrum
(\ref{Aj}) of a spherical surface:
\beq\label{Aal}
A =9.824\,l^2_p n = 6.678\,l^2_p N; \quad \alpha=6.678.
\eeq
As to the mass $M$ of a black hole, it is quantized in the units
of the Planck mass $m_p$ as follows:
\beq\label{M}
M^2= \,\frac{0.614}{\pi}\, m_p^2\, n.
\eeq

Let us present also for the sake of comparison the corresponding
results of \cite{asht}:
\[
A_a = 8\ln2\, l_p^2\, n =\,\frac{8\ln2}{\sqrt{3}}\,l_p^2\,N =3.202
l_p^2\,N; \quad \alpha_a=3.202; \quad M^2_a= \,\frac{\ln
2}{2\pi}\, m_p^2\, n =\,\frac{0.347}{\pi}\, m_p^2\, n\,.
\]

Of course, the solution proposed in \cite{asht} looks at least
more simple and elegant. On the other hand, the advantage of our
solution is that it is based on a simple and natural physical
conjecture.

It should be emphasized that in both cases one arrives at the
quantization rule for the black hole entropy (and area) with
integer quantum numbers $n$, as proposed in \cite{bek}.

In conclusion, let us comment briefly upon some previous
investigations of the considered problem. In \cite{rove} the
entropy is defined as logarithm of the number of microstates for
which the sum (\ref{N}) is between $N$ and $N+\Delta N$, $N\gg
\Delta N \gg 1$ (here and below the notations of the present
article are used). The conclusion made in \cite{rove} is that the
value of this logarithm is in the interval $(0.96\,-\,1.38)N$.
However, under the only condition $N \gg 1$, without any
assumption made about the distribution of the angular momenta $j$
over the edges, how can one arrive at the above numbers
$(0.96\,-\,1.38)N$? The same question (in fact, objection) refers
to the results obtained in \cite{kra}.

\bigskip
\bigskip
\begin{center}***\end{center}
I am grateful to G.G. Kirilin, I.V. Kolokolov, V.V. Sokolov, and
especially to Ya.I. Kogan and R.V. Korkin for the interest to the
work and useful discussions. The work was supported in part by the
Russian Foundation for Basic Research through Grant No.
01-02-16898, through Grant No. 00-15-96811 for Leading Scientific
Schools, by the Ministry of Education Grant No. E00-3.3-148, and
by the Federal Program Integration-2001.

\end{document}